\documentstyle[preprint,aps,epsfig]{revtex}

\newcommand{\beq}{\begin{eqnarray}}
\newcommand{\eeq}{\end{eqnarray}}
\newcommand{\beqs}{\begin{eqnarray}}
\newcommand{\eeqs}{\end{eqnarray}}
\newcommand{\bary}{\begin{array}}
\newcommand{\eary}{\end{array}}

\newcommand{\figpos}{p}        

\newcommand{\Vev}[1]{\left\langle #1\right\rangle}

\def\eps{\epsilon}

\def\plb#1{Phys.\ Lett.\ {\bf B#1}}

\def\prd#1{Phys.\ Rev.\ {\bf D#1}}
\def\prl#1{Phys.\ Rev.\ Lett. {\bf#1}}

\def\epjc#1{Eur.~Phys.~J.\ {\bf C\,#1}}

\def\gsim{\ \rlap{\raise 3pt \hbox{$>$}}{\lower 3pt \hbox{$\sim$}}\ }
\def\lsim{\ \rlap{\raise 3pt \hbox{$<$}}{\lower 3pt \hbox{$\sim$}}\ }
\def\overset#1#2{\ \rlap{\raise 3pt \hbox{$#1$}}{\lower 1pt\hbox{$#2$}}\ }
\def\underset#1#2{\ \rlap{\lower 3pt \hbox{$#1$}}{\raise 2pt \hbox{$#2$}}\ }
\def\eq#1{{\rm eq.}~(\ref{#1})}
\def\Eq#1{{\rm Eq.}~(\ref{#1})}

\def\putFig#1#2#3#4#5#6#7 
{
 \newpage
 \begin{figure}[\figpos]
 \LARGE
 $#7$
 \begin{center}
 \mbox{\epsfig{figure=#1.eps,angle=0,width=#3cm,height=#4cm}}
 $#6$
 \end{center}
 \vspace{1.5cm}
 \caption{#2}
 \label{#1}
 \end{figure}
}

\begin{document}

\preprint{\vbox{\hbox{WIS/19/00-SEPT-DPP}
                \hbox{hep-ph/0009280}}}

\title{~Lepton Masses and Mixing Angles with \\
         Spontaneously Broken O(3) Flavor Symmetry} 

\author{Gilad Perez \\ 
        \small \it Department of Particle Physics,
        Weizmann Institute of Science,
        Rehovot 76100, Israel}
\maketitle

\begin{abstract}
We present a model
based on an O(3) flavor symmetry and a minimal extension 
of the scalar sector to induce hierarchical breaking,
${\rm O(3)}\rightarrow{\rm O(2)}
\rightarrow{\rm SO(2)}
\rightarrow\,$nothing.
The model naturally accounts for all the known lepton parameters
and yields various interesting predictions for others:
(i) Neutrinos are nearly degenerate, $m_\nu\sim0.1$ eV;
(ii) The solar neutrino problem is solved by the MSW small 
     mixing angle solution;
(iii) The MNS mixing angle $\theta_{13}$ is unobservably small,
      $\theta_{13}={\cal O}(10^{-5})$.

\end{abstract}
\vspace{1cm}

\section{Introduction}

The SuperKamiokande atmospheric neutrino experiment
\cite{atm} has provided evidence for neutrino masses.
There have been many attempts to understand the lepton flavor
pattern in the framework of flavor symmetries.
Models with an O(3) or SO(3) flavor symmetry~\cite{O3} 
are particularly interesting
due to the fact that the O(3) symmetry naturally 
accommodates degenerate neutrinos. 
The possibility of nearly degenerate neutrinos 
is in agreement with all the 
experimental data and is motivated by the possibility that
neutrinos play a role in the evolution of the 
large-scale structure of the universe (see {\it e.g.}~\cite{astro} 
and references therein).

In this work we present an economical model based on a spontaneously 
broken O(3) flavor symmetry, which
naturally satisfies the constraints from the experimental data and
yields several non-trivial predictions for upcoming neutrino 
experiments. 

In order to make our discussion concrete we 
briefly review the experimental data
related to the lepton sector.
In an effective two neutrino framework, 
the data from the atmospheric neutrino experiments 
leads to the following results
(at 99$\%$ CL)~\cite{Gonzalez}:
\beq
\Delta m^2_{23}\sim 
(1-8)\times {10}^{-3}~ 
\rm{eV^2}
\,, \ \
\sin^22\theta_{23}> 0.85
\label{th23}
\,.
\eeq
In a two generation framework the solar neutrino 
experimental data \cite{Solar} 
yield several viable solutions for the mass difference and 
mixing angle~\cite{Gonzalez} (at 99$\%$ CL):
\begin{itemize}
\item[i.]
The large mixing angle (LMA) MSW~\cite{MSW} solution:
\beq 
\Delta m^2_{12}\sim 
10^{-5}-10^{-3}~\rm{eV^2}\,, \ \
\tan^2\theta_{12} \sim 0.1-1
\label{LMA}\,.
\eeq
\item[ii.]
The large mixing angle with a low mass-squared difference (LOW) and 
the quasi-vacuum oscillation (QVO) solution:
\beq
\Delta m^2_{12}\sim 
5\times 10^{-10}-3\times 10^{-7}~\rm{eV^2}\,, \ \
\tan^2\theta_{12} \sim0.4-3
\label{QVO}\,.
\eeq
\item[iii.]
The small mixing angle (SMA) MSW solution:
\beq 
\Delta m^2_{12}\sim 
(3-10)\times 10^{-6}~\rm{eV^2}\,, \ \ 
\tan^2\theta_{12} \sim 2\times 10^{-4}-2\times 10^{-3}
\label{SMA}\,.
\eeq
\end{itemize}
Combining the CHOOZ experiment results \cite{CHOOZ} 
with the solar and the atmospheric neutrino experiments 
yields the following constraints, in a three generation framework:
\beq
\Delta m^2_{23}\sim 
(1-7)\times 10^{-3}~\rm{eV^2}\,, \ \ 
\sin^2\theta_{13} < 0.075
\label{CHOOZ}\,.
\eeq
The experimental data\cite{DBD} 
from neutrinoless double beta decay indicates that (at 90$\%$ CL):
\begin{equation}
(M_\nu)_{ee}< 0.2~{\rm eV}
\label{dbd}\,,
\end{equation}
where $(M_\nu)_{ee}$ is the value of the (11) element
of the neutrinos mass matrix in the basis where both the
charged lepton mass matrix and the weak interaction couplings
are diagonal.
Finally the charged lepton masses are given by~\cite{PDG}:
\beq
m_e\simeq0.51~{\rm MeV}\,, \ \
m_\mu\simeq105.7~{\rm MeV}\,,\ \
m_\tau\simeq1777~{\rm MeV}\,.
\label{mcl}
\eeq

\section{The Model}

We consider an effective theory with a cut-off scale $M$.
We include nonrenormalizable terms 
induced by the integration out of the heavy degrees of freedom,
with masses larger than $M$.  

The Lagrangian is invariant under an O(3) flavor symmetry.
The field content consist of the SM fields and additional
SM gauge singlet scalar fields.
The additional fields are
$S^{ij}$, $i,j\in\{1..3\}$, a traceless symmetric field 
which transforms as a $\bf 5$ of the
O(3) symmetry;
$\Phi^j$, a triplet of the O(3) symmetry
and $A$, a pseudo-singlet of the O(3) symmetry.
The lepton SU(2)$_L$ doublets, $L^i$, transform as a triplet of the flavor group. 
Two of the right handed charged leptons,  
$E_R^2$ and $E_R^3$,  
are singlets of the O(3)
symmetry, while the third one, $E_R^1$,
is a pseudo-singlet.

The masses and the mixing angles of the leptons are
induced by the nonzero VEVs of the scalar fields.
We assume that CP is conserved in the lepton sector.
Therefore the couplings in the Lagrangian and the 
VEVs of the scalars are real.
The flavor symmetry is broken by three small parameters,
$\delta_{1,2}$ and $\eps$,
as follows:
\beq
{\rm O(3)}\,
\underset{\ \ \ _{\delta_1}}
{\overset{^{\Vev{S^{ij}}}}{\,\longrightarrow}}\, {\rm O(2)}\,
\underset{\ \ \ _{\delta_2}}
{\overset{^{\,\,\Vev{A}}}{\longrightarrow}\, {\rm SO(2)}}\,
\underset{\ \, \, _{\eps\delta_1}}
{\overset{^{\Vev{\Phi^i}}}{\longrightarrow}}\,\emptyset
\,,
\label{break}
\eeq
with
\beq 
{\Vev{S^{ij}}\over M}&=&\delta_1\cdot diag(1,1,-2)\,, \ \
{\Vev{A}\over M}=\delta_2 \,, \ \
{\Vev{\Phi^i}\over M}=\eps\delta_1\cdot(0,\sin\alpha,\cos\alpha)
\,.
\eeq

\subsection{The Neutrino Mass Matrix}\label{Mnu}

Neutrino masses are related to the following terms in the Lagrangian:
\begin{equation}
{\cal L}_\nu= \left\{L^i L^i+\frac{1}{M} a L^iL^jS^{ij}
+\frac{1}{M^2}[ b (L^i\Phi^i)^2+ b' L^j L^j \Phi^i\Phi^i]\right\}
{HH\over M}+h.c.
\label{LmO3} \, ,
\end{equation}
where $H$ is the Higgs field
and the coefficients $a,b,b'$ are of order unity.
Given the breaking pattern of eq.~(\ref{break}), 
the Lagrangian in eq.~(\ref{LmO3}) 
induces the following neutrino mass matrix:
\begin{eqnarray}
M_\nu &=& m\pmatrix
 {
  1+ a \delta_1+ b' \delta_1^2\eps^2&0&0\cr
  0&1+ a \delta_1+ b' \delta_1^2\eps^2+ b \delta_1^2\eps^2 s^2&
  b\delta_1^2\eps^2 sc\cr
  0&b\delta_1^2\eps^2   sc&
  1-2 a \delta_1+b'\delta_1^2\eps^2+b\delta_1^2\eps^2 c^2
 }.\nonumber\\
\label{MO3cd33}
\end{eqnarray}
with $m={\Vev{H}^2\over M}$, $s=\sin\alpha$ and $c=\cos\alpha$.

The mixing angles $\theta^{\nu}_{ij}$ required to diagonalize 
the mass matrix in eq. (\ref{MO3cd33}) are given by:
\beq
\tan2\theta^{\nu}_{23}&\approx&2{\delta_1\eps^2  b sc\over 
                       3 a +\delta_1\eps^2  b (s^2-c^2)}\ll1\,,
\nonumber\\
\theta^{\nu}_{13}\,,\theta^{\nu}_{12}&\approx&0
\label{genmixn44} \, .
\eeq
Subleading corrections to eq.~(\ref{genmixn44}) appear
when higher dimension operators such as
$\eps^{ijk}AL^m\Phi^m \Phi^n S^{ni}\Phi^j L^k{HH\over M^6}$
are added to the Lagrangian in eq.~(\ref{LmO3}).

The eigenvalues of (\ref{MO3cd33}) are given, to leading order, 
by: 
\beq
m_{\nu_1} &=&m(1+a\delta_1+ b'\delta_1^2\eps^2)\,, 
\nonumber\\
m_{\nu_2} &=&m(1+a\delta_1+ b'\delta_1^2\eps^2+b\delta_1^2\eps^2 s^2)
\,,
\nonumber\\
m_{\nu_3} &=&m(1-2a\delta_1+b'\delta_1^2\eps^2+b\delta_1^2\eps^2 c^2)
\,.
\label{sev}
\eeq
Since the contributions to the mixing angles from the neutrino
sector in eq.~(\ref{genmixn44}) 
are negligibly small, the mixing angles of the
MNS matrix~\cite{MNS} will be determined by the
charged lepton mass matrix.

\subsection{Charged Leptons Mass Matrix}\label{Mcl}

Charged lepton masses are related to the following terms in the 
Lagrangian:
\beq
{\cal L}_{\ell}&=&\left\{\left(a_3{L^i}^\dagger\Phi^i
+b_3\frac{1}{M}{L^i}^\dagger S^{ij}\Phi^j\right)E_R^3+
\left( a_2{L^i}^\dagger\Phi^i
+b_2\frac{1}{M}{L^i}^\dagger S^{ij}\Phi^j\right)E_R^2\right.
\nonumber\\
&+&\left.
{A\over M}\left[\left( a_1{L^i}^\dagger\Phi^i
+b_1\frac{1}{M}{L^i}^\dagger S^{ij}\Phi^j\right)E_R^1
+d_3\frac{1}{M^2}\epsilon^{ijk}{L^i}^\dagger\Phi^l S^{lj}\Phi^k E_R^3
\right.\right.
\nonumber\\
&+&\left.\left.
d_2\frac{1}{M^2}\epsilon^{ijk}{L^i}^\dagger\Phi^l S^{lj}\Phi^k E_R^2\right]
+
d_1\frac{1}{M^2}\epsilon^{ijk}{L^i}^\dagger\Phi^l S^{lj}\Phi^k E_R^1
\right\}\frac{H}{M}+h.c.
\label{RSO44}\,,
\eeq 
where the coefficients $a_i,b_i$ and $d_i$ are of order unity.
Given (\ref{RSO44}) and the breaking pattern of eq.~(\ref{break}) 
the following mass matrix is obtained:
\beq
M_{\ell}&=&m_{\ell}\pmatrix{3 d_1sc\delta_1^2\eps&
                      3 d_2 sc\delta_2\delta_1^2\eps&
                      3 d_3 sc\delta_2\delta_1^2\eps\cr
                      s\delta_2( a_1+ b_1\delta_1)&
                      s( a_2+ b_2\delta_1)&
                      s(a_3+ b_3\delta_1)\cr
                      c\delta_2( a_1-2 b_1\delta_1)&
                      c( a_2-2 b_2\delta_1)&
                      c(a_3-2 b_3\delta_1)
                     } 
\label{Matcl} \, ,
\eeq
where $m_{\ell}=\langle H\rangle\delta_1\eps$.

As we saw, the mixing angles in the 
neutrino sector are negligibly small (\ref{genmixn44}).
Therefore, to leading order, the mixing angles required 
to diagonalize the charged lepton
mass matrix in eq.~(\ref{Matcl}) will determine the 
mixing angles of the MNS matrix.
Thus the angle $\theta_{23}$ of the MNS matrix
is given by:
\begin{eqnarray}
\tan2\theta_{23}
\approx-\tan 2\alpha+{\cal O}\left(\delta_1,\delta_2^2\right)
\label{MNS23} \, .
\eeq
It is of order unity, in agreement with the experimental data 
[eq.~(\ref{th23})].
The mixing angle $\theta_{13}$
is given by:
\begin{eqnarray}
\tan2\theta_{13}&=&{\cal O}\left(\eps\delta_2\delta_1^2\right)\ll 1
\label{MNS13} \,,
\eeq
in agreement with eq.~(\ref{CHOOZ}).
The angle $\theta_{12}$ is:
\begin{eqnarray}
\tan2\theta_{12}&=&{\cal O}(\eps\delta_2)\ll 1
\label{MNS12} \, .
\eeq
The eiganvalues of (\ref{Matcl}) are given, to leading order, by
\beq
m_e &=& 3m_{\ell}  d_1\delta_1^2\eps sc
\,, 
\nonumber\\
m_\mu &=&3m_{\ell} \delta_1 sc
{(a_3 b_2+ a_2 b_3)\over a_{23}}
\,,
\nonumber\\
m_\tau &=&m_{\ell}a_{23}\,,
\label{ev}
\eeq
where $a_{23}\equiv\sqrt{a_2^2+ a_3^2}$.
The relations in eq.~(\ref{ev}) are valid as long as
$\delta_2\lsim \delta_1$.


\subsection{Consistency and Predictions}

In the previous subsections~(\ref{Mnu}) and (\ref{Mcl})
we found expressions for the
six masses and three mixing angles - the flavor
parameters of the lepton sector.
Using the experimetal data given in the first section,
we can constrain the four free parameters 
$\delta_{1,2},\eps,M$,
test the model and identify its predictions
for the upcoming experiments. 

\subsubsection{Constraining the Small Parameters} 
The charged lepton masses are known rather accurately, consequently
they provide stringents constraint on the free parameters.
The known ratio between the muon and the tau masses 
[\eq{mcl}] and the corresponding predicted ratio 
given in eq.~(\ref{ev}) give: 
\beq
{m_\mu \over m_\tau}\approx 3\delta_1 sc
{a_3 b_2+ a_2 b_3\over a_{23}^2}\ \ \Longrightarrow \ \
\delta_1={\cal O}\left(0.03\right)
\label{del1}\,.
\eeq
From eqs.~(\ref{mcl}), (\ref{ev}) and (\ref{del1}) we further find:
\beq
{m_\tau\over\langle H\rangle}
\approx\delta_1\eps a_{23}\ \ \Longrightarrow \  \
\eps={\cal O}\left(0.3\right)
\label{eps}\,.
\eeq
In order to set the allowed range of $\delta_2$ and $M$
we turn to the neutrino sector.\\
\Eq{MNS12} implies that the only possible solution to the solar neutrino 
problem in our model is the SMA solution.
Using eqs.  (\ref{SMA}), (\ref{MNS12}) and (\ref{eps})
we get:
\beq
\tan\theta_{12}={\cal O}\left(\eps\delta_2\right)
\ \ \Longrightarrow \ \
\delta_2={\cal O}\left(0.1-0.01\right)
\label{del2}\,.
\eeq
Using eqs. (\ref{SMA}), (\ref{sev}), (\ref{del1}) and (\ref{eps}) we get:
\beq
\Delta m^2_{12}\sim 2bs^2 m^2\delta_1^2\eps^2\ \ \Longrightarrow \ \
m={\cal O}\left(0.1~{\rm eV}\right)\ \ \Longrightarrow \ \
M={\cal O}\left(10^{14}~{\rm GeV}\right)
\label{M}\,.
\eeq
Note that the large scale of $M$,
or equivalently of $\Vev{S^{ij}}$ and $\Vev{\phi^i}$, makes
any  process related to the correponding
massless Goldstone bosons practically unobservable~\cite{FY}.  


\subsubsection{Consistency Checks} 
The lepton sector contains nine CP conserving flavor parameters,
constrained by the experimental data as given
in eq.~(\ref{th23})
and eqs.~(\ref{SMA})-(\ref{mcl}).
Four of them were used to find the values of the model free parameters.
This implies that there are five more relations that
can be compared with the corresponding
experimental data and either test the model or give predictions.
We have already pointed that the mixing angles $\theta_{23}$
and $\theta_{13}$
given in eqs. (\ref{MNS23}) and (\ref{MNS13})
satisfy the constraints in eqs. (\ref{th23}) and (\ref{CHOOZ}).
There are three more non-trivial consistency
checks:
\begin{itemize}
\item[(i)] The well known ratio between the electron and muon masses:
\beq
{m_e \over m_\mu}
\approx\delta_1\eps{d_1 a_{23}\over a_3 b_2+ a_2 b_3}
={\cal O}\left(10^{-2}\right)
\label{rcl12}\,,
\eeq
is consistent with \eq{mcl}.
\item[(ii)] The neutrinos mass-squared difference between the second and
third generation:
\beq
\Delta m_{23}^2\approx 6m^2 a\delta_1
= {\cal O}\left(10^{-3}~{\rm eV^2}\right)
\label{rn23}
\,,
\eeq
is consistent with \eq{th23}.
\item[(iii)] The bound from neutrinoless double beta decay 
translates into a constraint on $(M_\nu)_{ee}$, defined below \eq{dbd}.
In our case it is well approximated by
$(M_\nu)_{11}$ which was calculated in \eq{MO3cd33}. Hence:
\beq
(M_\nu)_{11}\sim m={\Vev{H}^2\over M}= 
{\cal O}\left(10^{-1}~{\rm eV}\right)
\label{Mee}
\,,
\eeq
is consistent with \eq{dbd}.
\end{itemize}
To explicitly demonstrate the phenomenological consistency of our model,
we set numerical values 
(the numbers are not necessarily the most favorable ones)
to $\delta_i,\eps$ and $M$:
\beq
\delta_1=0.035\,,\ \,
\delta_2=0.03\,,\, \
\eps=0.26\,,\,\
M=1.5\cdot10^{14}~{\rm GeV}\,.\label{num}
\eeq
Substituting the numerical values for the above parameters yields
the following values for the different observables:
\beq
\left[{m_e\over 0.51 ~{\rm Mev}}\right] &\simeq& 1.5\cdot d_1 \sin2\alpha
\,,\ \,
\left[{m_\mu\over 106 ~{\rm Mev}}\right] \simeq 0.8\cdot
{a_3 b_2+ a_2 b_3 \over a_{23}}\sin2\alpha
\,, \nonumber \\
\left[{m_\tau\over 1780 ~{\rm Mev}}\right]&\simeq& 0.9\cdot a_{23}\,;
\nonumber \\
(M_\nu)_{ee}&\sim&m\simeq0.2~{\rm eV}\,,\ \,
{\Delta m^2_{12}}\simeq 7\cdot 10^{-6} \cdot b \sin^2\alpha~{\rm eV^2}\,,
\nonumber \\
{\Delta m^2_{23}}&\simeq& 9\cdot10^{-3} \cdot a ~{\rm eV^2}
\,,\ \,\tan\theta_{12}={\cal O}(\eps\delta_2)\sim 10^{-2}
\label{numsub}
\,.
\eeq
Taking into account the unknown coefficient of order one,
\eq{numsub} fit reasonably well the experimental data.

\subsubsection{Predictions}
The model contains several non trivial predictions:
\begin{itemize}
\item The neutrinos are nearly degenerate with mass $m$, 
$m={\cal O}\left(0.1~{\rm eV}\right)$.
\item The correct solution of the solar problem is the SMA, with
      the ratio between the square mass difference $\Delta m^2_{12}$ and
      $\Delta m^2_{23}$ given by:
       \beq
         {\Delta m^2_{12}\over\Delta m_{23}^2}
          ={\cal O}\left(10^{-3}\right)  
         \label{rn1223}.
       \eeq
\item The ratio between the small mixing angles $\theta_{13}$ 
       and $\theta_{12}$ is given by:
\beq
{\theta_{13}\over\theta_{12}}={\cal O}\left(10^{-3}\right)\,.
\eeq
\end{itemize}


\section{Summary and colclusion}

We presented a model
with an O(3) flavor symmetry that is  spontaneously broken by
hierarchical VEVs of scalars,
${\rm O(3)}\rightarrow{\rm O(2)}\,
\rightarrow {\rm SO(2)}
\rightarrow\,$nothing.
The lepton flavor parameters were derived  
from the most general Lagrangian consistent 
with the symmetry and taking
all dimensionless couplings to be of order unity.
The model naturally accounts for all the known lepton parameters,
passes several consistency checks
and yields various interesting predictions:
(i) Neutrinos are nearly degenerate, $m_\nu\sim0.1$ eV.
(ii) The solar neutrino problem is solved by the MSW small 
     mixing angle solution. 
(iii) The MNS mixing angle $\theta_{13}$ is unobservably small,
      $\theta_{13}={\cal O}(10^{-5})$.

\acknowledgements
 
I thank Yossi Nir, Sven Bergmann and Oleg Khasanov
for helpful discussions and comments on the
manuscript.


{}


\begin{thebibliography}{99}

\bibitem{atm} Y. Fukuda {\it et al.}, Phys. Lett. {\bf B433}, 9 (1998)
[hep-ex/9803006]; 
{\bf B436}, 33 (1998) [hep-ex/9805006]; 
Phys. Rev. Lett. {\bf 81}, 1562 (1998) [hep-ex/9807003];
{\bf 82}, 2644 (1999) [hep-ex/9812014]. 


\bibitem{O3}
See {\it e.g.}
R. Barbieri, L. J. Hall, G. L. Kane and G. G. Ross, hep-ph/9901228;
C. Wetterich, Phys. Lett. {\bf B451}, 397 (1999) [hep-ph/9812426];
C. D. Carone and M. Sher, Phys. Lett. {\bf B420}, 83 (1998)
   [hep-ph/9711259];
A. Ghosal, hep-ph/9905470;
M. Tanimoto, T. Watari, T. Yanagida, Phys. Lett. {\bf B461}, 345 (1999)
[hep-ph/9904338];
Y-L. Wu,  Int. J. Mod. Phys. {\bf A14}, 4313 (1999) [hep-ph/9901320];
Nucl. Phys. Proc. Suppl. {\bf 85}, 193 (2000) [hep-ph/9908436]; 
Eur. Phys. J. {\bf C10}, 491 (1999) [hep-ph/9901245]; 
\prd{60}, 073010 (1999) [hep-ph/9810491]; 
E. Ma, \plb{456}, 48 (1999) [hep-ph/9812344];
P. Bamert, C.P. Burgess, \plb{329}, 289 (1994) [hep-ph/9402229];
A.S. Joshipura, Z. Phys. {\bf C64}, 31 (1994).


\bibitem{astro}
J.R. Primack and M. A. K. Gross, Proceedings 
of the Xth Rencontres de Blois, "The Birth of Galaxies", 
28 June - 4, July, 1998 [astro-ph/9810204];
J.R. Primack, J. Holtzman, A. Klypin and D.O. Caldwell, 
Phys. Rev. Lett. {\bf 74}, 2160 (1995) [astro-ph/9411020].


\bibitem{Gonzalez}M.C. Gonzalez-Garcia, parallel talk in
ICHEP  2000 (Osaka, Japan, July 28, 2000);
M.C. Gonzalez-Garcia and C. Pena-Garay,
hep-ph/0009041.

\bibitem{Solar} R. Davis, Prog. Part. Nucl. Phys. {\bf32}, 13 (1994); 
Y. Fukuda {\it et al.}, Phys. Rev. Lett. {\bf77}, 1683 (1996);
P. Anselmann {\it et al.}, Phys. Lett. {\bf B357}, 237 (1995); 
{\bf B361}, 235 (1996).

\bibitem{MSW} L. Wolfenstein, 
Phys. Rev. {\bf D17}, 2369 (1978); S.P. Mikheev and A.Y. Smirnov,
Sov. J. Nucl. Phys.  {\bf 42}, 913 (1985); 
Nuovo Cim.  {\bf C9}, 17 (1986).

\bibitem{CHOOZ}M. Apollonio {\it et al.}, Phys. Lett. {\bf B420}, 397 (1998)
[hep-ex/9711002].


\bibitem{DBD}L. Baudis {\it et al.}, 
Phys. Rev. Lett. {\bf83}, 41 (1999) [hep-ex/9902014].  

\bibitem{PDG}D.E. Groom {\it et al.}, 
\epjc{15}, 1 (2000). 

\bibitem{MNS}Z. Maki, M. Nakagawa and S. Sakata, Prog. Theo. Phys. 
{\textbf 28}, 870 (1962).


\bibitem{FY}M. Fukugita and T. Yanagida, \prl{55}, 2645 (1985);
Y. Chikashige, R. N. Mohapatra, R. D. Peccei, Phys. Lett. {\bf 98B},
265 (1981);
D. B. Reiss, Phys. Lett. {\bf 115B}, 217 (1982); 
F. Wilczek, Phys. Rev. Lett. {\bf 49}, 1549 (1982); 
G. Gelmini, S. Nussinov and T. Yanagida, Nucl. Phys. {\bf B219}, 
31 (1983);
D.S. Dearborn, D.N. Schramm and G. Steigman,
\prl{56}, 26 (1986);
J.L. Feng {\it et. al.}, \prd{57}, 5875 (1998) [hep-ph/9709411].

\end{thebibliography}
\end{document}